\documentclass[3p,times,procedia]{elsarticle}
\usepackage{nupha_ecrc}

\volume{00}

\firstpage{1}

\journalname{Nuclear Physics A}

\runauth{Q}

\jid{nupha}

\jnltitlelogo{Nuclear Physics A}

\usepackage{amssymb,amsmath,graphicx,mathrsfs}

\usepackage[figuresright]{rotating}
\usepackage[utf8]{inputenc}

\begin{document}

\begin{frontmatter}



\dochead{XXVIIth International Conference on Ultrarelativistic Nucleus-Nucleus Collisions\\ (Quark Matter 2018)}

\title{A resummed method of moments for the relativistic hydrodynamic expansion}


\author[a,b]{L. Tinti}

\author[a]{G. Vujanovic}

\author[c]{J. Noronha}

\author[a,d,e]{U. Heinz}

\address[a]{Department of Physics, The Ohio State University, Columbus, Ohio 43210, USA}
\address[b]{Institut f\"ur Theoretische Physik, Johann Wolfgang Goethe-Universit\"at, Max-von-Laue-Str.~1, D-60438 Frankfurt am Main, Germany}
\address[c]{Instituto de F\'isica, Universidade de S\~ao Paulo, S\~ao Paulo 05508-090, Brazil}
\address[d]{Theoretical Physics Department, CERN, CH-1211 Gen\`eve 23, Switzerland}
\address[e]{ExtreMe Matter Institute (EMMI), GSI Helmholtzzentrum f\"ur Schwerionenforschung, 
                Planckstrasse 1, D-64291 Darmstadt, Germany}

\begin{abstract}

The relativistic method of moments is one of the most successful approaches to extract second order viscous hydrodynamics from a kinetic underlying background. The equations can be systematically improved to higher order, and they have already shown a fast convergence to the kinetic results. In order to generalize the method we introduced long range effects in the form of effective (medium dependent) masses and gauge (coherent) fields. The most straightforward generalization of the hydrodynamic expansion is problematic, or simply ill defined, at higher order. Instead of introducing an additional set of approximations, we propose to rewrite the series in terms of moments resumming the contributions of infinite non-hydrodynamics modes. The resulting equations are are consistent with hydrodynamics and well defined at all order. We tested the new approximation against the exact solutions of the Maxwll-Boltzmann-Vlasov equations in $(0+1)$-dimensions, finding a fast and stable convergence to the exact results.

\end{abstract}


\end{frontmatter}


\section{Introduction}
\label{sect:Intro}

Relativistic hydrodynamics has been extensively used to describe the properties of the hot, dense matter produced in relativistic-heavy-ion collisions~\cite{Heinz:2013th}. Hydrodynamics appears to be predictive, even if the usual assumptions of small gradients and small deviations from equilibrium are routinely violated in the early stages of evolution~\cite{Schenke:2012wb, Schenke:2017bog,Bozek:2016kpf}. In fact, there is a mounting evidence that the gradient expansion is actually divergent, both in the strong~\cite{Buchel:2016cbj} and weak coupling regime~\cite{Denicol:2016bjh}. This divergent behavior does not imply that the lower orders of the expansion can't be used in practice (see for instance~\cite{Romatschke:2017ejr}, and citations therein), but it precludes to use the higher orders to improve the approximation or to estimate if hydrodynamics can be applied or not. The slow roll expansion shows a similar divergent behavior~\cite{Denicol:2017lxn}, however the method of moments is free from such divergences (also shown in Ref.~\cite{Denicol:2016bjh}). The most important shortcoming of the method of moments is the microscopic background; typically the relativistic Botlzmann equation for a single particle species~\cite{Denicol:2014loa}. It is therefore important to generalize the hydrodynamic expansion from the method of moments to a physically more general state. We considered mixed gasses with a Vlasov term coupling the (quasi)particles to coherent fields. The final aim would be, though, to generalize to the full relativistic quantum case, namely using substituting the distribution function with the more fundamental Wigner quasi-probability functions~\cite{DeGroot:1980dk}.

\section{Method of moments and the resummed expansion}
\label{sect:main}

Many modern approaches to extract the evolution of macroscopic quantities from a kinetic substrate can be summarized as follow. Starting from the relativistic Boltzmann equation it is possible to write the (exact) time evolution of the distribution function

\begin{equation}\label{exact}
 p\cdot\partial f(x,p) =- {\cal C}[f] \Rightarrow \dot f = u\cdot \partial f = \frac{(p\cdot u)(u\cdot\partial f)}{(p\cdot u)}= -\frac{1}{(p\cdot u)}\left[ \vphantom{\frac{}{}}{\cal C} -p\cdot\nabla f \right].
\end{equation}
The dot stands for the comoving derivative $u\cdot \partial$, $\nabla_\mu=\Delta^\alpha_\mu\partial_\alpha$ for the spatial gradient, and $\Delta^{\mu\nu}= g^{\mu\nu} -u^\mu u^\nu$ the projector orthogonal to the four-velocity $u^\mu$. Thanks to Eq.~(\ref{exact}) it is possible to write the exact evolution of any moment of the distribution function, in particular of the tensors of the form

\begin{equation}\label{reducible}
 {\cal F}^{\mu_1\cdots \mu_l}_r = \int_{\bf p} (p\cdot u) p^{\mu_1}\cdots p^{\mu_l} f,
\end{equation}
being $\int_{\bf p}=\int d^3 p/E_{\bf p}$  the Lorentz-covariant momentum integral. In particular,  familiar stress-energy tensor is $T^{\mu\nu}={\cal F}_0^{\mu\nu}$. Making use of Eq.~(\ref{exact}), after some algebra one obtains

\begin{equation}\label{moments_ev}
 \dot{\cal F}^{\mu_1\cdots \mu_l}_r + {C}^{\mu_1\cdots \mu_l}_{r-1} = r \dot u_\alpha {\cal F}^{\alpha\mu_1\cdots \mu_l}_{r-1} - \nabla_\alpha{\cal F}^{\alpha\mu_1\cdots \mu_l}_{r-1} +(r-1)\nabla_\alpha u_\beta {\cal F}^{\alpha\beta\mu_1\cdots \mu_l}_{r-2},
\end{equation}
the tensors ${C}^{\mu_1\cdots \mu_l}_r$ are defined following the same prescription as in~(\ref{reducible}), except with the collisional kernel instead of the distribution function. In particular, for the $r=0$, $l=2$ case

\begin{equation}\label{Tmunu_ev}
 \dot{T}^{\mu\nu} + {C}^{\mu\nu}_{-1} =  - \nabla_\alpha{\cal F}^{\alpha\mu\nu}_{-1} -\nabla_\alpha u_\beta {\cal F}^{\alpha\beta\mu\nu}_{-2}.
\end{equation}
It can be proved exactly that the time projection of the last equation is the local conservation of energy and momentum, namely $\partial_\mu T^{\mu\nu}=0$. The other equations correspond to the exact evolution of the pressure correction. Equations~(\ref{moments_ev}), couples alway to moments of different rank $s$ and energy index $r$. The moments defined in~(\ref{reducible}) are not independent ($u_\alpha {\cal F}^{\alpha \beta_1\cdots}_r = {\cal F}_{r+1}^{\beta_1\cdots}$) but there is always a coupling to an independent component in of the higher ranking tensor, making the set of equations infinite. There are more prescriptions to extract hydrodynamics from Eq.~(\ref{Tmunu_ev}), they all differ, essentially, on the way to treat the leftover independent components of the higher ranking tensors as sown in detail in Ref.~\cite{Denicol:2014loa}. In fact, of the modern approaches to anisotropic hydrodynamics fall in this category too~\cite{Tinti:2015xwa,Molnar:2016vvu}. It is not necessary to consider only the lower order approximations, as it has been done in ref.~\cite{Denicol:2016bjh}. Including systematically the approximated degrees of freedom as new variables and using Eq.~(\ref{moments_ev}) for the higher ranking moments as the dynamical evolution.

\begin{figure*}[t]
\begin{center}
\includegraphics[angle=0,width=0.49\linewidth]{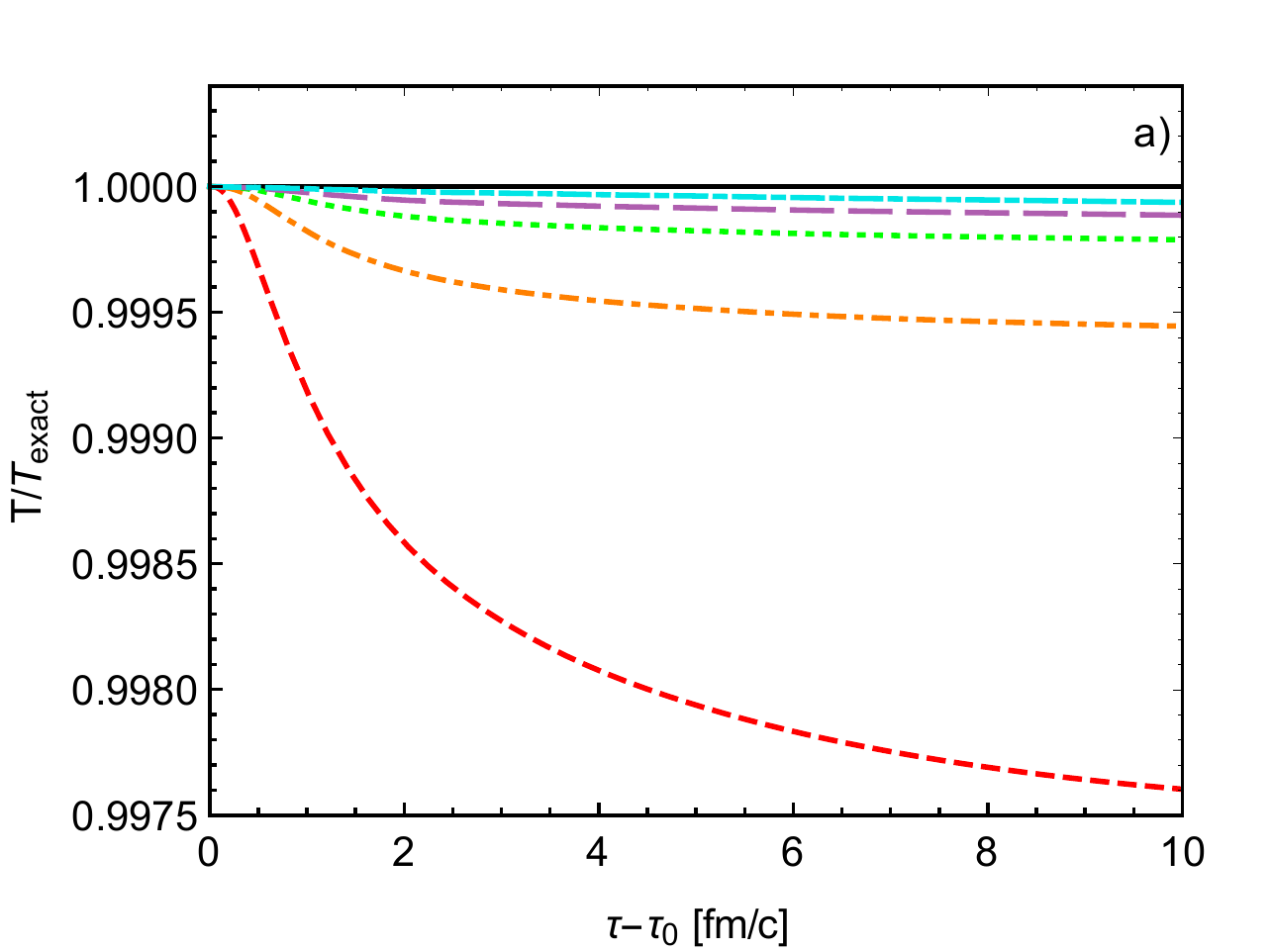}
\includegraphics[angle=0,width=0.49\linewidth]{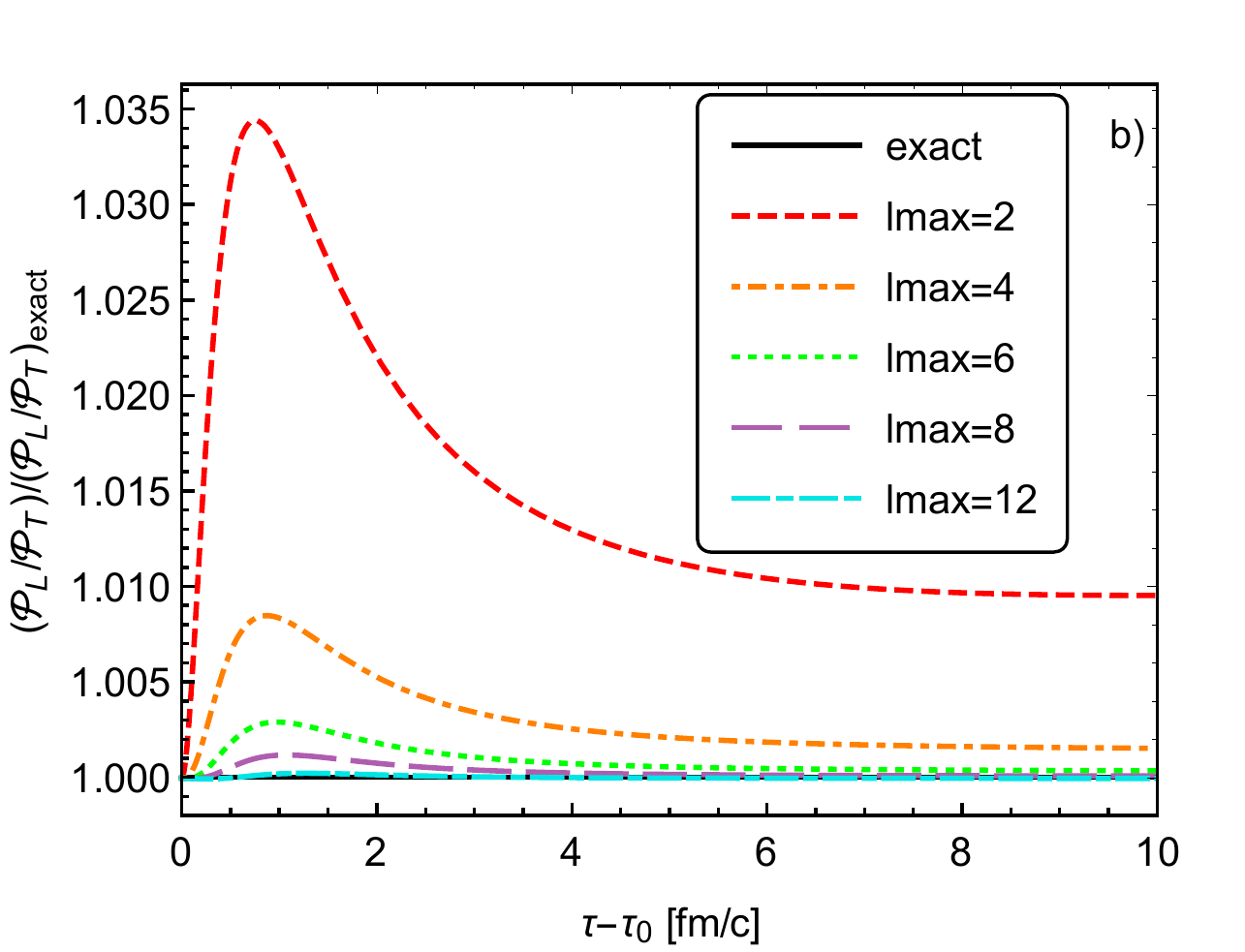}\\
\end{center}
\caption{(Color online) Relative precision of the approximation. The black solid lines correspond to the exact solution, the red, dashed one the leading order approximation, the orange dot-dashed NLO, the green dotted NNLO, the last two lines correspond, respectively, to dynamical tensors up to the 8th rank (purple, long-dashed) and 12th (light blue, long dot-dashed).
	}
\label{F1}
\end{figure*}

In the case of the Botzmann-Vlasov equation

\begin{equation}
\label{RBVE}
   p\cdot\partial f + m (\partial_\rho m) \, \partial^\rho_{(p)} f + q F_{\alpha\beta}p^\beta\partial^\alpha_{(p)} f = -{\cal C}[f],
\end{equation}
one would be tempted to use the same approach and use the updated equations for the moments in the case of long range interactions

\begin{equation}\label{BV_moments_ev}
\begin{split}
 \dot{\cal F}^{\mu_1\cdots \mu_l}_r + {C}^{\mu_1\cdots \mu_l}_{r-1} = & r \dot u_\alpha {\cal F}^{\alpha\mu_1\cdots \mu_l}_{r-1} - \nabla_\alpha{\cal F}^{\alpha\mu_1\cdots \mu_l}_{r-1} +(r-1)\nabla_\alpha u_\beta {\cal F}^{\alpha\mu_1\cdots \mu_l}_{r-2} + m\dot m \, (r-1) \, {\cal F}_{r-2}^{{\mu_1\cdots \mu_l}} \\ 
 & + s \; m\partial^{(\mu_1}m \, {\cal F}_{r-1}^{ {\mu_2\cdots \mu_l})} -q (r-1)\,  E_\alpha \, {\cal F}_{r-2}^{\alpha {\mu_1\cdots \mu_l}} -q\, s\, g_{\alpha\beta} F^{\alpha (\mu_1} {\cal F}_{r-1}^{ {\mu_2\cdots \mu_l} ) \beta}.
\end{split}
\end{equation}
In fact the first orders in the expansion are not problematic, but the coupling with the the non-dimensionless mass and electro-magnetic terms implies that some moments have a lower physical dimension than the other ones. Order by order one would add integrals with lower energy index $r$ as new degrees of freedom. After a division by the appropriate power of the temperature (in order to have a dimensionless quantity) and after a change of variables ${\bf p} = m {\bf y}$ these integral read
\begin{equation}
\label{suggests}
   \frac{{\cal F}_r^{\mu_1\cdots \mu_l}}{T^{r+l+2}} = \left( \frac{m}{T} \right)^{r+l+2}
   \int_0^\infty dy \left(\sqrt{1{+} y^2}\right)^{r-1} y^{l+2}\int d\Omega\, \hat{y}^{\mu_1}\cdots  \hat{y}^{\mu_l} f(x,m {\bf y}).
\end{equation}
In the ultra-relativistic limit $m/T{\,\to\,}0$, the moments diverge with $r{+}l<-2$ diverge; for the equilibrium distribution this is immediate to check. One way to proceed would be to use a non dynamical approximation for the $r<0$ moments, as suggested in~\cite{Denicol:2012cn}. We prefer instead to regularize this infrared problem making use of a different set of moments (resummed moments)

\begin{equation}\label{Gen1}
 \Phi^{\mu_1\cdots \mu_l}(x,\xi^2) = \int_{\bf p} (p\cdot u) p^{\mu_1}\cdots p^{\mu_l} e^{-\xi^2(p\cdot u)^2} f(x,p), 
\end{equation}
which re-sum the contribution of infinitely many ${\cal F}$ moments, as one can see expanding the Gaussian term inside the integrand. All the (non-resummed) moments can be recovered from the new set

\begin{equation}\label{Gen_to_F}
{\cal F}^{\mu_1\cdots \mu_l}_{2n} = \lim_{\xi\to 0} (-\partial_{\xi^2})^n\Phi^{\mu_1\cdots \mu_l}, \qquad {\cal F}^{\mu_1\cdots \mu_l}_{2n-1} = \frac{2}{\sqrt{\pi}}\int_0^\infty d\xi (-\partial_{\xi^2})^n\Phi^{\mu_1\cdots \mu_l}.
\end{equation}
The exact evolution of the resummed moments reads

\begin{eqnarray}\label{Gen_mom_ev_1}
 && \dot \Phi^{\mu_1\cdots \mu_l} +\delta \Phi_{\rm coll.}^{\mu_1\cdots \mu_l} =  -2\xi^2 \left[ \vphantom{\frac{}{}} \partial_\alpha u_\beta \, \Phi^{\alpha\beta\mu_1\cdots \mu_l} + m\dot m \, \Phi^{\mu_1\cdots \mu_l} - q E_\alpha \, \Phi^{\alpha\mu_1\cdots \mu_l} \right] \\
 &&+\frac{2}{\sqrt{\pi}}\int_\xi^\infty \!\!\!\! d\zeta  \frac{\zeta}{\sqrt{\zeta^2 -\xi ^2}}\left\{ \vphantom{\frac{}{}} \dot u_\alpha \, \Phi^{\alpha {\mu_1\cdots \mu_l}} -\nabla_\alpha \Phi^{\alpha {\mu_1\cdots \mu_l}} \vphantom{\frac{}{}} + s \left[ m\partial^{(\mu_1}m \, \Phi^{ {\mu_2\cdots \mu_l})} -q\, g_{\alpha\beta} F^{\alpha (\mu_1} \Phi^{ {\mu_2\cdots \mu_l} ) \beta} \right]\right\} . \nonumber
\end{eqnarray}
Differentiating and integrating by $\xi$ one recovers all the Eqs.~(\ref{BV_moments_ev}) with well-defined moments, recovering second order viscous hydrodynamics at the leading order (dynamical tensor up to rank two) and extending it  at the next orders (maximum rank increased by two). We tested the approximation against the exact solution of the simplest non-trivial case of the Boltzmann-Vlasov equation in $0+1$-dimensions. Namely, a gas of mass-less particles-antiparticles, with the collisional kernel treated in relaxation time approximation (RTA), expanding in a longitudinal boost invariant and transverse homogeneous fashion within a longitudinal electric field evolving according to the Maxwell equations. The symmetry of the expansion requires a vanishing net charge (hence the need of a mixed system) and a longitudinal-only electric field. We use the method described in Ref.~\cite{Ryblewski:2013eja} the exact solutions. It is possible to see in figure~(\ref{F1}) that the resummed moment expansion converges fast to the exact results.

%
%

\section{Conclusions}
\label{sect:out}

The method of moments for the relativistic hydrodynamic expansion can be generalized in an efficient way if one uses set of resummed moments. The resulting equations recover explicitly second order viscous hydrodynamics at the leading order, and can be systematically improved to higher order without numerical problems. An important remark is that there is no assumption of small gradients or small deviations from equilibrium at all orders. The expansion seems to converge mainly because most of the coupling to the higher ranking tensors appears canonically with a $-2\xi^2$ multiplying factor, in fact all the non-trivial ones (linearly independent from the lower ranking moments) in  the $0+1$-dimensional expansion. The Gaussian factor in the definition of the resummed moments~(\ref{Gen1}) mean that they are significantly different from zero only in the small $\xi$ region, making the coupling to higher order moments a small correction. It is therefore important to check if it is possible to extend further the formalism to the ful quantum case (Wigner quasiprobability function instead of distribution functions) and see if it is possible to justify the success of second order viscous hydrodynamics in extreme, very far from equilibrium, situations.

\section*{Acknowledgement}
This work was supported in part by the U.S. Department of Energy (DOE), Office of Science, Office for Nuclear Physics under Award No. \rm{DE-SC0004286}, the Collaborative Research Center CRC-TR 211 ``Strong-interaction matter under extreme conditions'' funded by DFG, and by the Fulbright Program.





\bibliographystyle{elsarticle-num}
\bibliography{Tinti_Proceedings_QM2018}







\end{document}